\documentclass[twocolumn]{openjournal}
\usepackage{natbib}
\usepackage{graphicx}
\usepackage{graphicx,amsmath,amssymb,amstext}
\usepackage{amsbsy,amsfonts,amsthm,color}
\usepackage{xcolor}
\usepackage[colorlinks,linkcolor=blue,citecolor=blue,urlcolor=blue ]{hyperref}

\newcommand{\half}{\frac{1}{2}}

\newcommand{\bomu}{\ensuremath{\boldsymbol{\mu}}}

\newcommand{\sfS}{\ensuremath{{\sf{S}}}}

\newcommand{\sfF}{\ensuremath{{\sf{F}}}}
\newcommand{\sfQ}{\ensuremath{{\sf{Q}}}}
\newcommand{\sfM}{\ensuremath{{\sf{M}}}}

\newcommand{\mathd}{\ensuremath{\mathrm{d}}}
\newcommand{\both}{\ensuremath{\boldsymbol{\theta}}}

\newcommand{\calP}{\ensuremath{\mathcal{P}}}

\newcommand{\calL}{\ensuremath{\mathcal{L}}}

\newcommand{\calG}{\ensuremath{\mathcal{G}}}

\newcommand{\calD}{\ensuremath{\mathcal{D}}}

\newcommand{\fatp}{\ensuremath{\boldsymbol{p}}}
\newcommand{\fatdelta}{\ensuremath{\boldsymbol{\Delta}}}

\newcommand{\sfL}{\ensuremath{{\sf{L}}}}

\newcommand{\vecw}{\ensuremath{{\vec{w}}}}
\newcommand{\vecW}{\ensuremath{{\vec{W}}}}

\newcommand{\sfC}{\ensuremath{{\sf{C}}}}
\newcommand{\invsfC}{\ensuremath{{\sf{C}^{-1}}}}

\usepackage{soul}

\usepackage{xcolor}

\begin{document}

\title{MCMC generation of cosmological fields far beyond Gaussianity}

\author{Joey R. Braspenning}
\affiliation{Leiden Observatory, Leiden University, Oort Gebouw, Niels Bohrweg 2, NL-2333 CA Leiden, The Netherlands.}

\author{Elena Sellentin}
\email{sellentin@strw.leidenuniv.nl}
\affiliation{Mathematical Institute, Leiden University, Snellius Gebouw, Niels Bohrweg 1, NL-2333 CA Leiden, The Netherlands.\\Leiden Observatory, Leiden University, Oort Gebouw, Niels Bohrweg 2, NL-2333 CA Leiden, The Netherlands.
}

\begin{abstract}
Structure formation in our Universe creates non-Gaussian random fields that will soon be observed over almost the entire sky by the Euclid satellite, the Vera-Rubin observatory, and the Square Kilometre Array. An unsolved problem is how to analyse best such non-Gaussian fields, e.g.~to infer the physical laws that created them. This problem could be solved if a parametric non-Gaussian sampling distribution for such fields were known, as this distribution could serve as likelihood during inference. We therefore create a sampling distribution for non-Gaussian random fields. Our approach is capable of handling \emph{strong} non-Gaussianity, while perturbative approaches such as the Edgeworth expansion cannot. To imitate cosmological structure formation, we enforce our fields to be (i) statistically isotropic, (ii) statistically homogeneous, and (iii) statistically independent at large distances. We generate such fields via a Monte Carlo Markov Chain technique and find that even strong non-Gaussianity is not necessarily visible to the human eye. We also find that sampled marginals for pixel pairs have an almost generic Gauss-like appearance, even if the joint distribution of all pixels is markedly non-Gaussian. This apparent Gaussianity is a consequence of the high dimensionality of random fields. We conclude that vast amounts of non-Gaussian information can be hidden in random fields that appear nearly Gaussian in simple tests, and that it would be short-sighted not to try and extract it.
\end{abstract}

\maketitle

\section{Introduction}
Random fields are ubiquitous in cosmological research, as they permeate the Universe ever since its earliest phases. In fact, the theory of cosmological inflation reasons that the initial distribution of matter in our Universe traces back to quantum fluctuations, and our Universe is thus inherently random from the outset \citep{Mukhanov,RuthStructureFormation}.

At the cosmic time of recombination, a snapshot of the statistical state of the Universe's matter fields was generated, which has been observed as the cosmic microwave background (CMB) by a series of experiments \citep{BICEP,WMAP9,3ACT,Planck2018}. These data reveal that the initial state of cosmic fields is compatible with Gaussian distributions whose power spectra are by now well measured.

Primordially generated non-Gaussianity is theoretically expected to be weak, but a detection of it would further constrain the physics of inflation. Accordingly, there exists a vast body of literature focusing on weakly non-Gaussian fields and their analysis, e.g.~\citet{Komatsu,Maldacena,Planck2018NG}.

In the post-CMB cosmic evolution, physical processes altered the statistical properties of the random cosmic fields, thereby creating structures which are richer than those of a Gaussian random field. Amongst these processes ranges gravity which -- unlike primordial mechanisms -- produces \emph{strong} non-Gaussianity, thereby immediately hindering a direct transfer of CMB analysis methodology to gravitationally evolved fields. In this paper, we hence focus on random fields which are strongly non-Gaussian. 

Besides gravity, other mechanisms generating strong non-Gaussianity include also the radiation of the first stars which burn bubble-like patterns into the cosmic hydrogen distribution. Non-Gaussian fields therefore also occur in studies of cosmic reionization \citep{JonathanI,JonathanII}. Current observations of non-Gaussian fields include e.g.~galaxy clustering and cosmic shear experiments \citep{DES, KiDS}. In the near future, the Euclid satellite \citep{Euclid}, the Vera-Rubin observatory \citep{LSST}, and the Square Kilometre Array (SKA) \citep{SKA} will supersede contemporary observations both in sky-coverage, depth, and data quality.

With these new observational capabilities, the window on non-Gaussian fields is opened. It is thus timely to focus on the statistical challenge of analyzing strongly non-Gaussian fields. This study complements a significant body of literature handling the classical differential equations that encode the physical laws which generate the non-Gaussianity. Such deterministic models of physics include full N-body simulations (e.g.~\citet{SpringelTNG}), and approximations including \citet{Zeldovich} and the second order Lagrangian perturbation theory (e.g. described in \citet{2LPT}). These approaches model the deterministic growth of non-Gaussian structures from random initial conditions. Once these deterministic codes output a final non-Gaussian field, a statistical solution is required to compare the modelled random fields to observed fields. There currently exists no consensus on how this comparison is achieved best, and in this paper we study it from a Bayesian point of view.

Targeting the inference of parameters from a non-Gaussian field in the long run, we begin by generating non-Gaussian fields from a flexible non-Gaussian sampling distribution. The reason for doing so, is that the sampling distribution becomes the likelihood when analyzing an observed field. We clarify this setup in Sect.~\ref{sec:segmentation}, and present our non-Gaussian ansatz, and fields generated from it, in sections \ref{sec:review}-\ref{sec:conclusions}.

\section{Inference from non-Gaussian Fields}
\label{sec:segmentation}
Increasingly accurate all-sky surveys confront contemporary cosmology with the challenge of inferring parameters from non-Guassian fields. In this section, we compare the problem to the more familiar case of inferring parameters from the Gaussian CMB, and we describe the challenges posed by the update to non-Gaussianity.

We imagine the initial data set is a pixelized random field, i.e. each pixel provides one data point and all pixels could be stored in a data vector. If $\Delta_i$ denotes the measured field value in the $i$th pixel, then the full data vector comprising all pixels will be $\fatdelta$. In case of the CMB one typically works with the spherical harmonic transform of $\fatdelta$, which are the usual modes $a_{\ell m}$. We collect all $a_{\ell m}$-modes in a data vector $\boldsymbol{a}_{\ell m}$. For fixed $\ell$, the variance of the $a_{\ell m}$-modes will be the spherical harmonic power spectrum $C_\ell$. The covariance matrix of $\boldsymbol{a}_{\ell m}$ will be diagonal for a full-sky observation, and we denote this covariance matrix by $\sfF = \mathrm{diag}(C_{\ell_{\rm min}},...,C_{\ell_{\rm max}})$. The unconventional choice of naming the CMBs covariance matrix $\sfF$ provides a slightly smoother transition to the upcoming non-Gaussianity study. 

Inferring physical parameters from the CMB implies the posterior probability of parameters $\both$ in light of the data $\boldsymbol{a}_{\ell m}$ must be computed. We shall denote this posterior distribution as $\calP(\both|\boldsymbol{a}_{\ell m})$.

Inference begins by specifying a prior $\pi(\both)$ on the physical parameters. In case of the CMB, drawing values for the parameters from the prior, the power spectra $C_\ell(\both)$ can be computed, implying in Bayesian jargon that the matrix $\sfF$ is `given' once the parameters $\both$ are fixed. 

Due to the CMB being a Gaussian random field, the $a_{\ell m}$-modes are drawn from a Gaussian sampling distribution of covariance matrix $\sfF$. The likelihood must therefore be the Gaussian distribution $\calG$, and the posterior simplifies to the familiar
\begin{equation}
     \calP(\both|\boldsymbol{a}_{\ell m})  \propto \calG\big[\boldsymbol{a}_{\ell m} | \sfF(\both)\big] \pi(\both).
\end{equation}
We now extend to inference with a non-Gaussian field. In this paper, we shall work in real space: instead of using $\boldsymbol{a}_{\ell m}$ we use the pixel values $\fatdelta$. For a non-Gaussian field, there will exist a matrix $\sfF$ as before, but we introduce two additional tensors $\sfS$ and $\sfQ$ that quantify the random field's non-Gaussianity, in a sense that will be described in due course.

Inference with a non-Gaussian field will again need to begin by specifying a prior $\pi(\both)$, but now all the tensors $\sfF(\both),\sfS(\both),\sfQ(\both)$ must be computed once $\both$ is given. The sampling distribution for non-Gaussian fields will then be $\calP(\fatdelta| \sfF,\sfS,\sfQ)$
and the posterior is then
\begin{equation}
     \calP(\both | \fatdelta)  \propto \calP\big[ \fatdelta | \sfF(\both), \sfS(\both) , \sfQ(\both) \big] \pi(\both).
\end{equation}
The probability distribution $\calP(\fatdelta| \sfF,\sfS,\sfQ)$ is the non-Gaussian sampling distribution of the fields, and the objective of this paper is hence to understand this probability distribution. The natural use of such a non-Gaussian distribution would be in a Monte-Carlo based cosmological inference, that ingests cosmic fields directly, see for example \citet{Murray, Taylor, Alsing}.

Providing a good ansatz for this non-Gaussian distribution is an intellectual challenge in itself, and precedes any numerical challenges. We thus devote Sect.~\ref{sec:review} to establishing a good ansatz, and in Sect.~\ref{sec:DALI_generate} we specialize our non-Gaussian distribution to statistically isotropic and homogeneous fields as in cosmology. In Sect.~\ref{section:pixelize} we discuss the numerical challenges of non-Gaussian field generation arising from the many permutations in non-Gaussian calculations and the high dimensionality of the problem. Example fields are then presented in Sect.~\ref{sec:results}.

\section{Ansatz for the Non-Gaussian Probability Distribution}
\label{sec:review}
In this section we describe our ansatz for a probability distribution from which non-Gaussian random fields can be drawn.

Describing non-Gaussianity in a simple mathematical form is challenging: per definition it is anything else but Gaussian, but no further information is provided. The standard approach to handling non-Gaussian fields uses higher-order moments or cumulants, which in the  case of studying random fields corresponds to computing 3-point and 4-point correlation functions, or even higher $n$-point correlation functions. Such cumulants can be used to construct the Edgeworth expansion, but this expansion neither yields positive-definite probability distributions, nor does it converge \citep{Edgeworth}. The hierarchy of correlation functions is accordingly known to capture the information of non-Gaussian fields only incompletely and inefficiently \citep{Carron}.

To overcome the problem of negative probabilities and the lack of convergence, \cite{DALI, DALII} (S14 and S15 from now on) therefore developed a novel series expansion of non-Gaussian distributions, called DALI. In comparison to the Edgeworth expansion, DALI provides a guaranteed positive distribution at all expansion orders. It inherits its convergence properties from a Taylor series, and has been shown to converge extremely rapidly: in practice, DALI reproduces extremely non-Gaussian distributions already after one or two terms beyond Gaussianity.

DALI, as published in S14 and S15, was developed as an extension of a Fisher matrix formalism. As a Fisher matrix approach does not include `real' data, we therefore begin by strongly modifying the original DALI setup: a data vector has to be introduced. The outcome of this reformulation will be that DALI is then able to analyze real data, or generate statistically compatible `fake' data.

\subsection{Review: DALI as a Fisher Matrix extension}

Here we briefly repeat the main points in the derivation of DALI as presented in S14 and S15 and derive the sampling distribution used in this paper.

If $d$ is the dimension of data space, and if $\both = (\theta_1,...,\theta_p)$ is a $p$-dimensional parameter vector to be inferred, then the original DALI-formulation provides an analytic approximation for the \emph{expected} non-Gaussian posterior of parameters, given Gaussian data with mean $\bomu(\both)$ and data covariance matrix $\sfC$. Hence, DALI approximates the posterior $\calP[\both| \bomu(\hat{\both})]$  meaning as in any Fisher forecasting framework, also the non-Gaussian DALI extension replaces not-yet taken data by their expectation value $\bomu(\hat{\both})$, evaluated at fiducial parameter values $\hat{\both}$ \citep{1997ApJ...480...22T, GenFish, 2016MNRAS.457.1490A}.

Denoting partial derivatives with respect to $\theta_\alpha$ by commas, $\partial_\alpha \bomu = \bomu,_\alpha$, the best fitting parameters by $\hat{\both}$, and the offset of the $\alpha$th parameter from its best fitting value as $\Delta_\alpha = \theta_\alpha -\hat{\theta}_\alpha$, we can Taylor expand the mean as
\begin{equation}
 \bomu(\both) = \bomu(\hat{\both}) +  \bomu,_\alpha \Delta_\alpha + \half  \bomu,_{\alpha\beta} \Delta_\alpha\Delta_\beta + ...,
\end{equation}
where Einstein's summation convention over repeated indices is implied. The DALI expansion of the posterior then follows to be
\begin{equation}
\begin{aligned}
     -\log \calP[\both|\bomu(\hat{\both})] &  = \frac{1}{2!} \left[ \bomu,_\alpha^\top \invsfC\bomu,_\beta\right]\Delta_\alpha \Delta_\beta\\
     & + \frac{1}{3!}\left[ 3 \bomu,^\top_{\alpha\beta} \invsfC \bomu,_\gamma\right]\Delta_\alpha \Delta_\beta \Delta_\gamma \\
     & + \frac{1}{4!} \left[ 3 \bomu,^\top_{\alpha\beta}\invsfC\bomu,_{\gamma\delta}\right] \Delta_\alpha \Delta_\beta \Delta_\gamma \Delta_\delta, \\
     & + ...
    \end{aligned}
       \label{mlogl}
\end{equation}
Here the factors $N!$ were not cancelled against the numerical prefactors in angular brackets, as the former are a reminder of the Taylor expansion from which Eq.~(\ref{mlogl}) originates, whereas the numerical prefactors arise from symmetries due to permuting derivatives and the symmetry of the scalar product. For details of this derivation, please refer to S14 and S15.

In Eq.~(\ref{mlogl}), we truncated the DALI expansion at second-order derivatives of the mean. For higher-order derivatives, further non-Gaussian distributions can be approximated, but already at second-order most single-peaked distributions are well approximated \citep{WBjoern}. In this paper, we thus focus on second-order expansions. The components of the tensors $\sfF, \sfS, \sfQ$ then read
\begin{equation}
\begin{aligned}
  F_{\alpha\beta} & =  \bomu,_\alpha^\top \invsfC\bomu,_\beta\\
  S_{\alpha\beta\gamma} & = 3 \bomu,^\top_{\alpha\beta} \invsfC \bomu,_\gamma \\
  Q_{ijkl} & = 3 \bomu,^\top_{\alpha\beta}\invsfC\bomu,_{\gamma\delta}.
    \end{aligned}
\end{equation}
We note that the tensors $\sfS$ and $\sfQ$ both contain second derivatives of $\bomu$ as inputs. They hence appear simultaneously in the expansion, and satisfy relations with respect to each other. This guarantees the positive definiteness of the DALI-expansion.

\section{DALI Field generation} 
\label{sec:DALI_generate}
In this paper we aim to use DALI to generate random fields under the constraints imposed by cosmology. In this section, we first modify the DALI distribution to meet this aim. We then introduce our pixelization scheme of the random field, and then present the implementation of the constraints arising from statistical isotropy and homogeneity due to the cosmological principle. We further impose independence at large distances due to causality and the finite speed of light, and domination of Gaussianity at large distances, as demanded by cosmology's theory of inflation, and the CMB.

We begin by modifying the DALI-distribution such that it can serve as sampling distribution for non-Gaussian fields. In other words, we want to draw $\fatdelta$ from the DALI distribution
\begin{equation}
    \fatdelta \sim \calP(\fatdelta| \sfF,\sfS,\sfQ).
\end{equation}
In this case, the distribution Eq.~(\ref{mlogl}) can be simplified:
The matrix $\invsfC$ in Eq. (\ref{mlogl}) plays the role of a metric, and in order for Eq.~(\ref{mlogl}) to be a positive definite function, the matrix $\invsfC$ has to be positive definite. It can therefore be Cholesky decomposed as
\begin{equation}
    \invsfC = \sfL \sfL^\top,
\end{equation}
where $\sfL$ is a unique lower triangular matrix, sometimes referred to as the `root'. Likewise, $\sfL^\top$ is upper triangular.

This allows us to introduce the vectors
\begin{equation}
\begin{aligned}
    \vecw_i & = \sfL^\top \bomu,_\alpha\\
    \vecW_{ij} & = \sfL^\top \bomu,_{\alpha\beta}.
    \end{aligned}
\end{equation}
Eq.~(\ref{mlogl}) can thus be rewritten as
\begin{equation}
\begin{aligned}
    -\mathrm{lnL}
     & = \frac{1}{2!} \left[ \vecw_i^\top \vecw_j\right]\Delta_i \Delta_j\\
     & + \frac{1}{3!}\left[ 3 \vecW^\top_{ij} \vecw_k\right]\Delta_i \Delta_j \Delta_k \\
     & + \frac{1}{4!} \left[ 3 \vecW^\top_{ij}\vecW_{kl}\right] \Delta_i \Delta_j \Delta_k \Delta_l \\
    \end{aligned}
    \label{mlogw}
\end{equation}
 We introduce three types of angles, $\vartheta,\eta,\zeta$ such that
\begin{equation}
\begin{aligned}
    \vecw_i^\top \vecw_j & = w_i w_j \cos(\vartheta_{ij}),\\
    \vecW^\top_{ij} \vecw_k & = W_{ij} w_k \cos(\eta_{ij,k} ),\\
    \vecW^\top_{ij}\vecW_{kl} & = W_{ij} W_{kl} \cos(\zeta_{ij,kl}),
    \end{aligned}
    \label{AngleDali}
\end{equation}
where $w_i, W_{ij}$ are the absolute values of $\vecw_i,\vecW_{ij}$.
To satisfy identity constraints, we set
\begin{equation}
\begin{aligned}
   \cos( \vartheta_{ii} ) & \equiv 1,\\
   \cos( \zeta_{ij,ij} ) & \equiv 1.
    \end{aligned}
\end{equation}
The remaining values, $w_i, W_{ij}, \cos(\theta_{ij}), \cos(\eta_{ij,k}), \cos(\zeta_{ij,kl})$ are free parameters that have to be set to satisfy physical constraints. For example, if all $W_{ij}$ were zero, a Gaussian random field would ensue, whose covariance matrix is defined by all $w_i$ and $\cos(\theta_{ij})$.

As the indices $i,j,k,l$ run over the pixels in the field, and as physics depends on distance, we now first have to introduce the pixelization scheme, before the values for $w_i, W_{ij}, \cos(\theta_{ij}), \cos(\eta_{ij,k}), \cos(\zeta_{ij,kl})$ can be further specified.

\subsection{Pixelization scheme} \label{section:pixelize}
The fields generated in this paper will be square and have N pixels.
We enumerate the pixels with roman indices $i \in [1,N]$. If $x(i)$ denotes the (Cartesian) x-coordinate of the $i$th pixel, and $y(i)$ its y-coordinate, then the distance $r$ between  two pixels $i$ and $j$ is given by
\begin{equation}
    r(i,j) = \sqrt{ \left(x(i) -x(j) \right)^2 + \left(y(i) - y(j)\right)^2  },
\end{equation}

The ordering of the pixels is irrelevant, all physics will depend only on r. 
In extension, the distance between a triplet or quadruplet of pixels is given by
\begin{align}
    &r(ij, k) = \sqrt{ r(i,k)^2 + r(j,k)^2},\\
    &r(ij, kl) = \sqrt{r(i,k)^2 + r(j,k)^2 + r(i,l)^2 + r(j,l)^2},
\end{align}
We will denote the vector of all pixel values as $\fatdelta$, the sidelength of the field $f = \sqrt{N}$. This means we will have N random variables in $\fatdelta$. Hence, field-generation quickly becomes a very high dimensional problem.

\begin{figure}
\centering 
\includegraphics[width=0.4\textwidth]{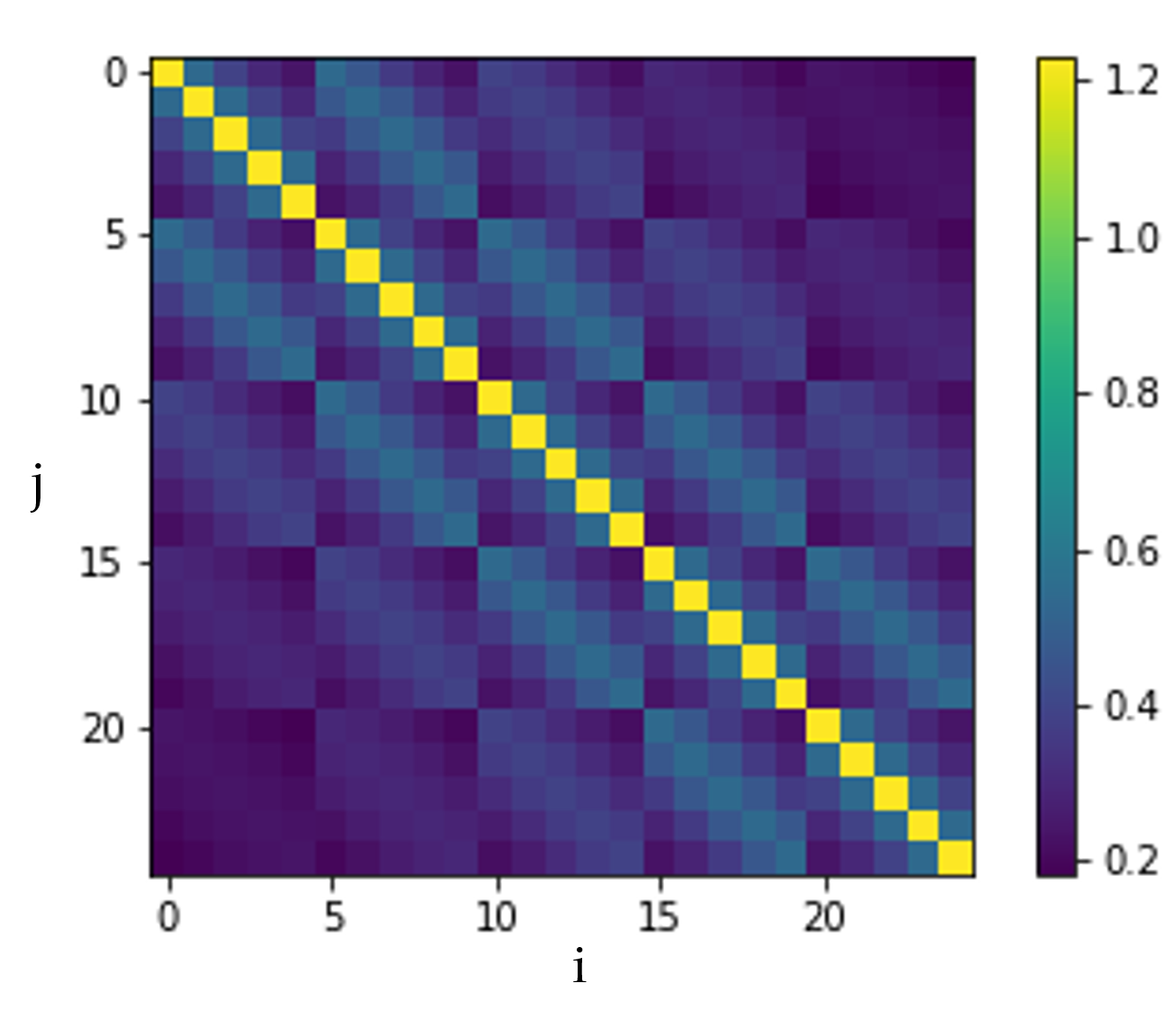}
\caption{Visualisation of the F tensor for a 25 pixel field. Colours indicate the value of $F_{ij}$ where $i$ and $j$ are pixel indices.}
\label{Ftensor}
\end{figure}   

\begin{figure}
\centering 
\includegraphics[width=0.4\textwidth]{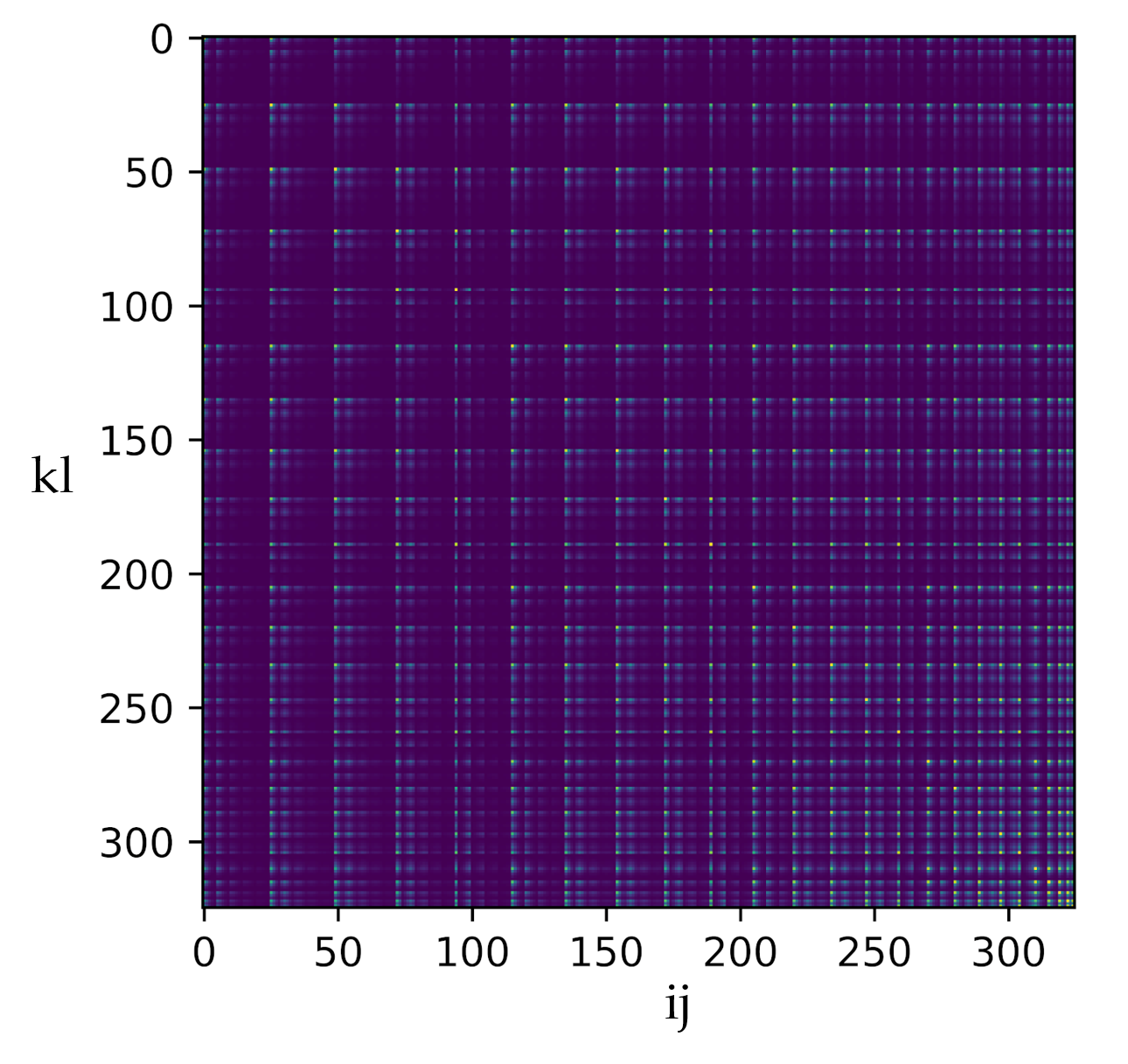}
\caption{Visualisation of the Q tensor for a 25 pixel field. Colours indicate the value of $Q_{ijkl}$, where $ij$ and $kl$ are unique pixel pairs. The visualisation is made for all such unique pixel pairs, of which there are 313 in a 25 pixel field.}
\label{Qtensor}
\end{figure}

\subsection{Statistical Isotropy and Homogeneity}
\label{sec:isohom}
On sufficiently large scales, standard cosmology assumes the Universe to be statistically homogeneous and isotropic. Statistical homogeneity is given by $f(\vec{x},\vec{x}') = f(\vec{x}-\vec{x}')$. Statistical isotropy then additionally updates this to $f(\vec{x}-\vec{x}') = f(|\vec{x}-\vec{x}'|)$. 

For future reference we note that the conditional distribution of density is no longer a simple function of distance on small scales. This implies that two point statistics are no longer sufficient to characterise a random field on such scales.
In this paper we focus on the largest cosmic scales, but our framework could be applied to smaller scales by modifying the tensors F, S and Q to make them dependent on more than only distance on small scales. An emergence of filamentary structures is hence not expected for the results here shown, but can (in future) be provided for.

\subsection{Independence and causality}
\label{sec:indep}
To ensure that different pixels of the field are statistically independent at large distances we impose the limit
\begin{equation}
    \cos(\theta_{ij}) \to 0 \ \mathrm{for \ } r(i,j) \to \infty
\end{equation}
which ensures that the covariance of widely separated pixels approaches zero.

Furthermore, to ensure that pixels which are far apart are causally disconnected we impose the following three constraints
\begin{equation}
\begin{aligned}
     W_{ij} & \to 0 \ \mathrm{for \ } r(i,j) \to \infty,\\
     \cos(\eta_{ij,k}) & \to 0 \ \mathrm{for \ } r(ij,k) \to \infty,\\
     \cos(\zeta_{ij,kl}) & \to 0 \ \mathrm{for \ } r(ij,kl) \to \infty,\\
    \end{aligned}
\end{equation}

\section{MCMC generation of non-Gaussian Fields} \label{sec:MCMC}

In this section we generate non-Gaussian random fields from DALI with a Monte Carlo Markov Chain (MCMC) approach.

As described in Sect.~\ref{section:pixelize}, we imagine the random field to be pixelated, as would occur when taking data. Each pixel $\Delta_i$ is then a random variable, and the collection of all pixels yields the random vector $\fatdelta$. The joint distribution of all pixel values is given by $\calD(\fatdelta|\sfF,\sfS,\sfQ)$ and defines which random patterns the field will form. 

A realization of one random field is then yielded by drawing from the distribution
\begin{equation}
    \fatdelta \sim \calD(\fatdelta|\sfF,\sfS,\sfQ).
    \label{draw}
\end{equation}
We therefore must succeed in creating an algorithm that indeed generates samples drawn as defined by Eq.~(\ref{draw}). The crux to succeeding will hereby be the high dimensionality of the distribution: for a square pixelized field of side length $f$, the number of random pixels to be jointly drawn will be $f^2$, therefore very rapidly encountering the `Curse of Dimensionality'.

We therefore generate our random fields with a \emph{Hamiltonian Monte Carlo} (HMC) sampler \citep{Betancourt}. Intuitively, a Hamiltonian Monte Carlo sampler can be understood as a clever way of increasing the distance between samples of a more conventional Metropolis-Hastings sampler. This is achieved by integrating  equations of motion to find a sample at similar probability. Readers not interested in the details of the HMC sampler may skip directly to the resulting fields presented in Sect.~\ref{sec:results}.

The Hamiltonian Monte Carlo algorithm begins by the observation that if a probability distribution is turned `upside down', then the former maximum will become the lowest point of a valley-like potential. The sampler then receives random kicks at selected points in time, and is left to propagate through the potential between the kicks. It thereby explores the probability distribution much more efficiently than a sampler with a Gaussian proposal, which has no guidance, centres on the old point and is generally very narrow and hence slowly step-by-step explore its local neighbourhood.

Implementing the HMC algorithm, we define the (negative) potential $\calL(\fatdelta)$ to be
\begin{equation}
   \calL(\fatdelta) = \log( \calD(\fatdelta|\sfF,\sfS,\sfQ) )
\end{equation}
We introduce momenta $\fatp$ and a mass matrix $\sfM$. The mass matrix is a tunable parameter that determines the efficiency of the sampler. The momenta $\fatp$ are auxiliary variables that are not of primary interest: they are introduced to upgrade the sampler to an evolving Hamiltonian system. The combination $(\fatdelta,\fatp)$ is accordingly referred to as `phase space'.

We define the Hamiltonian
\begin{equation}
    H(\fatdelta,\fatp) = - \calL(\fatdelta) + \half \log\left( |2\pi\sfM| \right) + \half\fatp^T\sfM^{-1} \fatp
\end{equation}
the augmented probability distribution is proportional to $e^{-H}$, implying
\begin{equation}
    \calD(\fatdelta|\sfF,\sfS,\sfQ) \to \calD(\fatdelta|\sfF,\sfS,\sfQ) \exp\left( -\half \fatp^T \sfM^{-1}\fatp\right),
\end{equation}
where we have ignored the normalisation coming from the mass matrix term as it is irrelevant for this work.
As the original distribution $\calD$ factors out, the probability distribution of $\fatdelta$ is the marginal of $e^{-H}$ over the momenta. This underlines that the momenta are indeed only auxiliary variables.

The HMC algorithm begins by selecting a starting point $\fatdelta_0$, and drawing an initial momentum vector
\begin{equation}
    \fatp_0 \sim \frac{1}{\sqrt{|2\pi \sfM|}}\exp\left(-\half \fatp^T \sfM^{-1} \fatp \right).
    \label{Drawp}
\end{equation}
The position and momentum are then updated by solving the Hamiltonian system
\begin{equation}
\begin{aligned}
\frac{\mathd \fatdelta}{\mathd t} & = \frac{\partial H}{\partial \fatp} = \sfM^{-1} \fatp,\\
\frac{\mathd \fatp}{\mathd t} & = - \frac{\partial H}{\partial \fatdelta} = \nabla_{\fatdelta} \calL(\fatdelta).
\end{aligned}
\label{eom}
\end{equation}
`Solving' here implies integrating these differential equations to track how the sampler propagates through phase space.

Numerically solving this system is prone to numerical inaccuracies, which will affect the acceptance rate of the sampler and might also break its reversibility. We hence choose the Leapfrog algorithm to solve the equations of motion Eq.~(\ref{eom}), which is a symplectic integrator conserving energy to second order yielding a trajectory with an energy that fluctuates around the true energy of the exact trajectory. The leapfrog algorithm is given by 
\begin{equation}
    \begin{aligned}
\fatp(t + \epsilon/2) & = \fatp(t) + \epsilon \nabla_{\fatdelta} \calL[\fatdelta(t)]/2,\\
\fatdelta(t + \epsilon) & = \fatdelta(t) + \epsilon \sfM^{-1} \fatp(t + \epsilon/2),\\
\fatp(t + \epsilon) & = \fatp(t + \epsilon/2) + \epsilon \nabla_{\fatdelta} \calL[ \fatdelta(t+ \epsilon) ]/2,
    \end{aligned}
    \label{LF}
\end{equation}
where $\epsilon$ is a small stepsize appearing due to having discretized time. The gradient and Hessian of a DALI distribution are given in App.~\ref{gradhess}. The three steps of Eq.~(\ref{LF}) are repeated $N$ times for $N$ updates of position. As $i \in [1,f^2]$ denotes the value $\Delta_i$ of each pixel, the leapfrog loops runs in parallel (and synchronized) for all pixels.

As the leapfrog algorithm's energy oscillates around the true energy of the integrated trajectory, stopping the integration at any point in time is likely to yield an energy somewhat different from that of an exact integration. We thus decide upon whether the end point $(\fatdelta_{\rm E},\fatp_{\rm E})$ is accepted as a valid sample via the acceptance criterion known from the Metropolis-Hastings algorithm. 
\begin{equation}
    P_{\rm accept} = {\rm min}\big[1, \exp\big(-H( \fatdelta_{\rm E},\fatp_{\rm E} ) - H(\fatdelta_0,\fatp_0)\big)\big].
\end{equation}
If the point $(\fatdelta_{\rm E},\fatp_{\rm E})$ is accepted, it takes the place of $\fatdelta_0$, otherwise $\fatdelta_0$ remains unchanged. Having completed these steps, the algorithm reruns from Eq.~(\ref{Drawp}) thereby building up a chain of samples.

\section{Results} \label{sec:results}
In this section we present results from a non-Gaussian field generated using the DALI distribution and an HMC sampler. We will use this non-Gaussian field to show the feasibility of generating such fields using DALI and an HMC sampler, and to demonstrate the subtleties involved in analysing them.

\begin{figure}
\centering 
\includegraphics[width=0.4\textwidth]{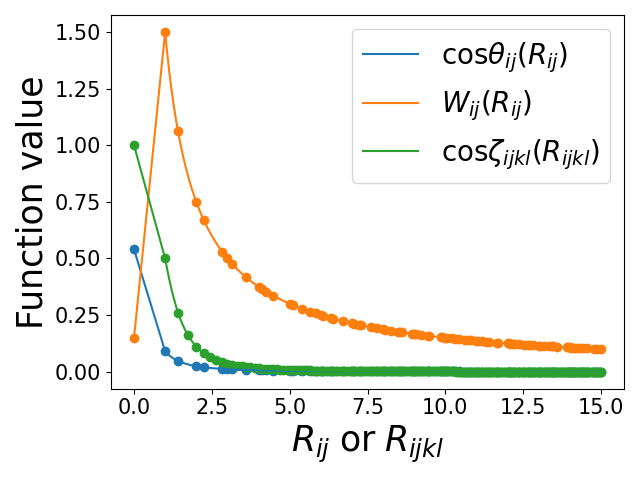}
\caption{Functional forms for the equations defined in \ref{PureNG}. Dots indicate the locations where these functions are evaluated.}
\label{equation26}
\end{figure}

The non-Gaussian field used in this section only has a Q tensor (see equation \ref{mlogl}) using
\begin{equation}
\begin{aligned}
w_i & = 0, \\
\cos(\eta_{ij,k}) & = 0,\\
    W_{ij} & =\frac{1.5}{r_{ij}},\\
    R & = \sqrt{r_{13}^2 + r_{14}^2 + r_{23}^2 + r_{24}^2}, \\
    \cos(\zeta_{ij,kl}) & = \frac{1}{R^3 + 1},\\
    \end{aligned}
    \label{PureNG}
\end{equation}
We will therefore call it a Q-field. Figure \ref{equation26} shows the distance dependence of these functions. Their decay towards zero at large pixel separations implies pixel far apart will be statistically independent. Homogeneity and isotropy are ensured by using functions of distances only.

We generate fields with side length $f =70$, which gives a total dimensionality of 4900. Sampling such non-Gaussian field is extremely expensive, and we therefore switch off non-Gaussianity beyond 1/8 th of the field's side length. This approximation is motivated by physical processes acting locally, with e.g. the electromagnetic and gravitational force diminishing as a function of distance and becoming causally disconnected at infitite distance.

Though our sampling is run using \texttt{C++} and makes use of parallelization on 128 CPUs as well as 128GB of memory, we are limited to a 70x70 pixel field. This is largely due to the many dot products in Eq. \ref{mlogw} which require retrieving stored pixel values a large number of times. We have extensively optimized the algorithm to make use of the isotropy and homogeneity of the field, both in storing the $W_{ij}W_{kl}$ values, the ordering of the for-loops and the combination of identical pixel combinations.

\begin{table}
    \centering
    \caption{Convergence diagnostics for the Gaussian and non-Gaussian chains.}
    \begin{tabular}{c|c|c}
        \toprule
        Type & Specifics & Value \\ 
        Gelman-Rubin & NG: 20 & 0.9989 \\
        Gelman-Rubin & NG: 10 & 0.9996\\
        Gelman-Rubin & NG: 2 & 1.0003\\
        Gelman-Rubin & G: 20 & 0.9990\\
        Gelman-Rubin & G: 10 & 0.9996\\
        Gelman-Rubin & G: 2 & 1.0003\\
        Correlation length & NG pixels & $2.17^{+0.026}_{-0.051}$\\
        Correlation length & NG lnL & 23.26 \\
        Correlation length & G pixels & $3.77^{+0.372}_{-0.371}$ \\
        Correlation length & G lnL & 34.90\\
    \end{tabular}
    \label{convtable}
\end{table}

\begin{figure*}
\centering 
\includegraphics[width=\textwidth]{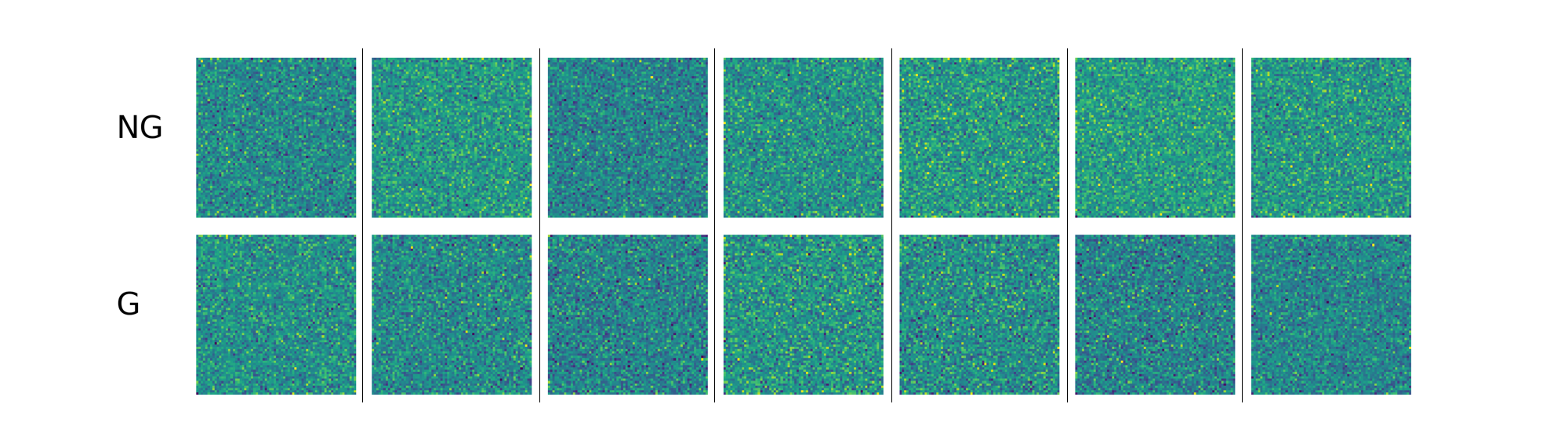}
\caption{Seven consecutive samples generated for a non-Gaussian Q-field (top) and its closest approximating Gaussian field which has the same covariance (bottom). This figure illustrates that the human eye struggles tell apart the Gaussian approximation and the original non-Gaussian field. Nonetheless, the non-Gaussian information is detected at high statistical significance in an analysis.}
\label{field_progression}
\end{figure*}  

One aim is to compare this non-Gaussian Q-field with a Gaussian field of equal covariance matrix. To this end, after generating the non-Gaussian field, we compute the sample covariance matrix from the MCMC chain and use it in the HMC sampler to generate a Gaussian field. The thus created Gaussian field can be interpreted as the closest Gaussian approximation to our non-Gaussian field.

As sampling in 4900 dimensions is difficult, we iteratively improved the quality of our chains by plotting the HMC phase space trajectory in all dimensions. We equally plotted the trajectory in pixel-space only, to judge which step sizes are adequate. With suitable step sizes found, we computed the correlation lengths \citep{corrlength} and the Gelman-Rubin convergence diagnostics \citep{GelmanRubin}. These are listed in Table~\ref{convtable} with NG abbreviating `non-Gaussian', `G' abbreviating `Gaussian', and followed by the number of chains from which the convergence diagnostics were computed. We conclude our chains decorrelate after 3 to 30 steps, and indeed converge to the target distribution.

Figure \ref{field_progression} shows seven consecutive samples from an arbitrary point in the chain for both the Gaussian and non-Gaussian field. From these images, it would be near impossible to tell them apart by eye. It illustrates that non-Gaussian fields can look deceivingly similar to Gaussian fields, even when their sampling distributions differ strongly.

\begin{figure*}
\centering 
\includegraphics[width=\textwidth]{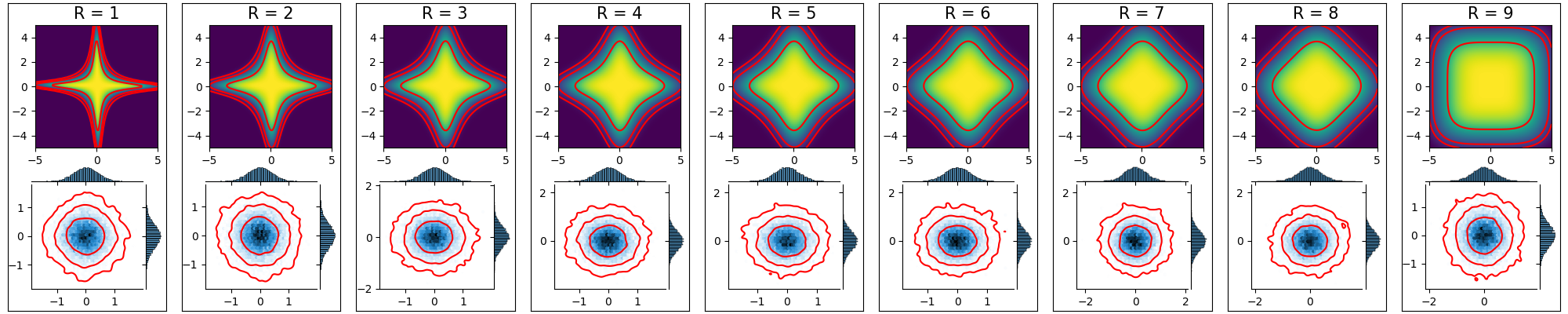}
\caption{Two dimensional conditional (top) and marginal (bottom) distributions for two pixels at 9 increasing distances $R$ (left to right). Going from left to right, the conditional joint distribution morphs from a star to a box, which is caused by it beginning to factorize for increasing distance. This transition was implemented to enforce causality, such that close pixels are co-dependent, while pixels with increasing separation become ever more independent (see Sect.~\ref{sec:indep}). The upper row (conditionals) shows clear non-Gaussian co-dependency between two pixels. The lower row imitates an an analysis of observed pixelized fields, where histograms of the pixel values are taken. It evidences that taking histograms of pixel values (marginalizing) easily triggers the false impression that the field were close to Gaussian.}
\label{conditionals_and_marginals}
\end{figure*}

The extreme non-Gaussianity of the Q-field's sampling distribution is illustrated by Figure \ref{conditionals_and_marginals}. The top row shows the conditionals of the non-Gaussian field, where the bottom row gives the marginals. The conditionals are clearly non-Gaussian. Figure \ref{conditionals_and_marginals} also shows that for small distances (left in the figure) between two pixels the non-Gaussian conditional cannot be factorized into to independent distributions for the two pixels. This implies close pixels are statistically co-dependent.  Towards the right of Figure \ref{conditionals_and_marginals} the distance between the studied pixels increases, and their joint probability distribution approaches a factorizable distribution. This is a visual impression of pixels at large distances becoming increasingly independent of each other.  

Figure \ref{conditionals_and_marginals} also illustrates that the strong non-Gaussianity of the sampling distribution is essentially invisible when plotting marginals of pixel pairs: the marginals in the lower row are all similar to an ellipsoidally contoured distribution. The washing out of the non-Gaussian shape of the sampling distribution is due to the high dimensionality of the sampling problem: the plotted marginals are integrals over 4898 dimensions and evidently the projection over many thousand dimensions will cause a loss of structure in the remaining two dimensions. Likewise, the two-dimensional conditionals in the top row only show such clear non-Gaussianity because all other dimensions are held fixed at zero when generating the conditional. When sampling, the probability of having a high dimensional system in such a state that all dimensions except two are exceedingly close to zero is negligible.

The appearance of shapeless marginals is particularly interesting in the context of how non-Gaussian fields are usually approached: often, histograms of pixel values are computed, and if these do not show a strong deviation from a Gaussian, then the field is judged as `close to Gaussian'. The Q-field here studied shows however that significant non-Gaussian information can be hidden in a field whose marginals look almost Gaussian. Accordingly, non-Gaussian analyses of cosmological fields should be attempted even when 1-point and 2-point analyses do not evidence any striking departure from Gaussianity.

\begin{figure}
\centering 
\includegraphics[width=0.4\textwidth]{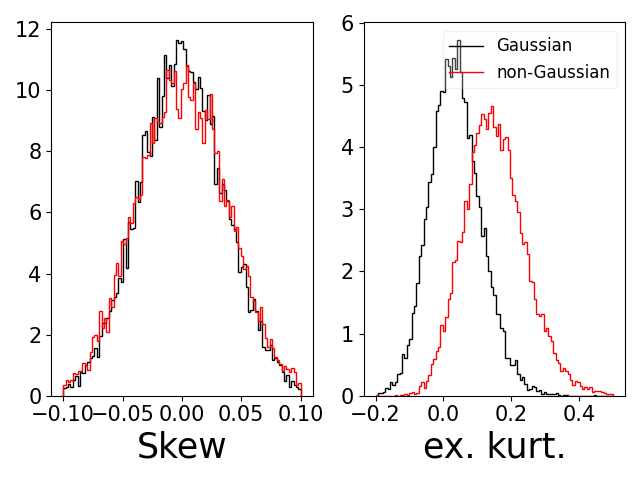}
\caption{Skew and excess kurtosis measured for different fields as sampled. A clear positive kurtosis can be observed for the non-Gaussian field.}
\label{skew_kurtosis}
\end{figure}

For example, we found a better discriminator between Gaussianity and non-Gaussianity to be the calculation of cumulants. We present measurements of the skewness and kurtosis in Figure \ref{skew_kurtosis}. The skewness is defined as
\begin{align}
    g = \frac{E\left[ (x - \Bar{x})^3 \right] }{\sigma^3} = \frac{\frac{1}{N}\sum_{i=0}^N (x_i - \Bar{x})^3}{\left[\frac{1}{N} \sum_{i = 0}^N (x_i - \Bar{x})^2\right]^{3/2}}.
\end{align}
Due to the use of the third power, the skewness measures asymmetry in a probability distribution. The Q-field has a symmetric distribution, due to the lack of a symmetry breaking $\sf{S}$ tensor. Accordingly, the skewness must be compatible with zero in our case, and non-Gaussianity is detected in the excess kurtosis, defined as
\begin{align}
    \kappa = \frac{E[(x - \Bar{x})^4]}{\sigma^4} = \frac{\frac{1}{N}\sum_{i=0}^N (x_i - \Bar{x})^4}{\left[\frac{1}{N} \sum_{i = 0}^N (x_i - \Bar{x})^2\right]^{2}}.
\end{align}
The skew and kurtosis are computed for each sampled field, with $N$ the number of pixels in the field. The resulting histograms for the skewness and curtosis are shown in Figure \ref{skew_kurtosis}. As expected, neither the Gaussian field, nor the Q-field show statistically significant departures of the skewness from zero. The kurtosis measurements show a clear difference between the Gaussian and non-Gaussian fields. While the kurtosis of the Gaussian fields is consistent with zero, the non-Gaussian fields show a positive offset with an average kurtosis of about 0.2.

We further note that all measurements of n-point functions where $n < 4$ show no disagreement between the Gaussian and non-Gaussian field. We verify isotropy and homogeneity of the field in App.~\ref{sec:verify_isohom}.

\begin{figure*}
 \centering
 \includegraphics[width=1\linewidth]{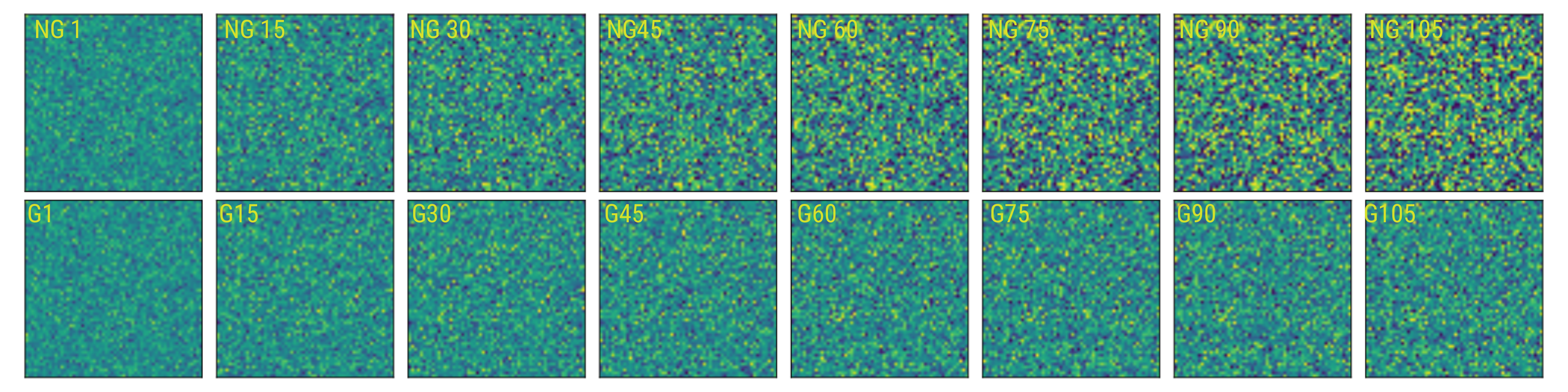}
 \caption{A non-Gaussian, and a Gaussian 60x60 field showing successive samples from the burn-in phase (left to right). Both fields have the same covariance matrix. As the sampler proceeds, the non-Gaussian field (top) builds up non-Gaussianities, and can be seen to quickly deviate from the best Gaussian approximation (bottom).} \vspace{16pt}
 \label{fig:sequence}
\end{figure*}

Finally, we study how an initially Gaussian field evolves randomly into a non-Gaussian field. To this end, we initialize the HMC sampler with a Gaussian field of mean zero and the same covariance matrix as the Q-field. We then let the sampler run in the potential well of the Q-field, such that the burn-in phase of the sampler causes a transition from Gaussianity to non-Gaussianity. Keeping the colour bar of the plotting range fixed then results in Figure \ref{fig:sequence}: We here show a 60x60 field, different from the 70x70 field shown before, that starts from a Gaussian and remains a Gaussian in the bottom row. Accordingly, the bottom row simply depicts consecutive samples of a Gaussian field from left to right. In contrast, the non-Gaussian Q-field is plotted in the top row, where the order left-to-right again shows consecutive samples: here we see that the initially Gaussian field transforms into a non-Gaussian field as a function of time. The emergence of more pronounced structures than in the Gaussian field can be seen by comparing the top and bottom row. Both fields start from the exact same initial configuration, but the non-Gaussian field can be seen to quickly deviate from this and form much stronger fluctuations in pixel values.

\section{Conclusions}
\label{sec:conclusions}
Non-Gaussian random fields are ubiquitous in observational cosmology. A commonly found wish is to compare an observed non-Gaussian field to a simulated one, in order to infer parameters via this statistical comparison. The ideal case for such a comparison would be if one knew the probability distribution that generated these fields: given this distribution, one could calculate how likely one field is, given the other.

In this paper we therefore developed a highly adaptable probability distribution from which non-Gaussian fields can be generated (Sect.~\ref{sec:DALI_generate}). In Sect. \ref{sec:isohom} we specialized to cosmology, by enforcing that our fields are statistically isotropic and homogeneous. We additionally enforced in Sect.~\ref{sec:indep} that pixels at increasing distance become statistically ever more independent of each other. 

Our field generation proceeds by MCMC sampling, and once the statistical properties of a field are defined by this distribution, our technique easily generates many thousands of fields with the prescribed statistics. However, the maximum resolution we could achieve with a university high performance cluster is 70 by 70 pixels. 

In fact, we found the numerical demands in generating the non-Gaussian fields to be unconventional and therefore (for now) limiting: drawing from a joint probability distribution of 4900 pixels implicitly implies all conditional distributions of one pixel's dependency on all others are to be evaluated. We therefore found the ideal numerical setup for this calculation to be a multi-thread access to an unconventionally large \emph{shared} memory, whereas more conventional distributed computing almost immediately produced long waiting times for accessing the required joint information.

As presented, our setup has the advantage of handling strong rather than perturbative non-Gaussianity. It is in this respect the first one to our knowledge. Although an extension to larger pixel numbers is highly desirable, we already conclude from our 70 by 70 fields that the human eye struggles to recognize non-Gaussianity, and also histograms of pixel values are an inaccurate predictor of how non-Gaussian a field is. The extraction of non-Gaussian information should therefore also be attempted on fields which appear almost Gaussian.

\appendix

\section{The gradient and Hessian}
\label{gradhess}
The gradient and Hessian of a DALI distribution appear as intermediate steps in the HMC sampler here used. Components-wise, the gradient is given by
\begin{equation}
\begin{aligned}
    \frac{\partial (- 2 \log(\calD)   )}{\partial \Delta_e}  & = 2 \vecw_e^\top\vecw_i \Delta_i + \vecw_e^\top \vecW_{ij}\Delta_i\Delta_j + 2 \vecW_{ei}^\top \vecw_j \Delta_i \Delta_j + \vecW_{ei}^\top \vecW_{jk}\Delta_i \Delta_j \Delta_k.
    \end{aligned}
    \label{gradient}
\end{equation}

The components of the Hessian matrix then follow to be
\begin{equation}
\begin{aligned}
    H_{ef} & = \frac{\partial^2 [- 2 \log(\calD)]}{\partial \Delta_e \partial \Delta_f} \\
   & =  2 \vecw_e^\top \vecw_f + ( \vecW_{ef}^\top \vecW_{ij} + 2 \vecW^\top_{ei}\vecW_{fj}) \Delta_i\Delta_j + 2 ( \vecw_e^\top \vecW_{fi} + \vecw_f^\top\vecW_{ei} + \vecw_i^\top\vecW_{ef}) \Delta_i. \\
    \end{aligned}
    \label{hessian}
\end{equation}
Evaluated at the peak, where $\fatdelta \equiv 0$, the Hessian is thus identical to the usual Fisher matrix. In cases of weak non-Gaussianity, it can provide a reliable estimator for the mass matrix of the sampler.
\vspace{12pt}

\section{Verification of statistical isotropy and homogeneity}
\label{sec:verify_isohom}
\begin{figure*}[h]
    \centering
    \includegraphics[width = \linewidth]{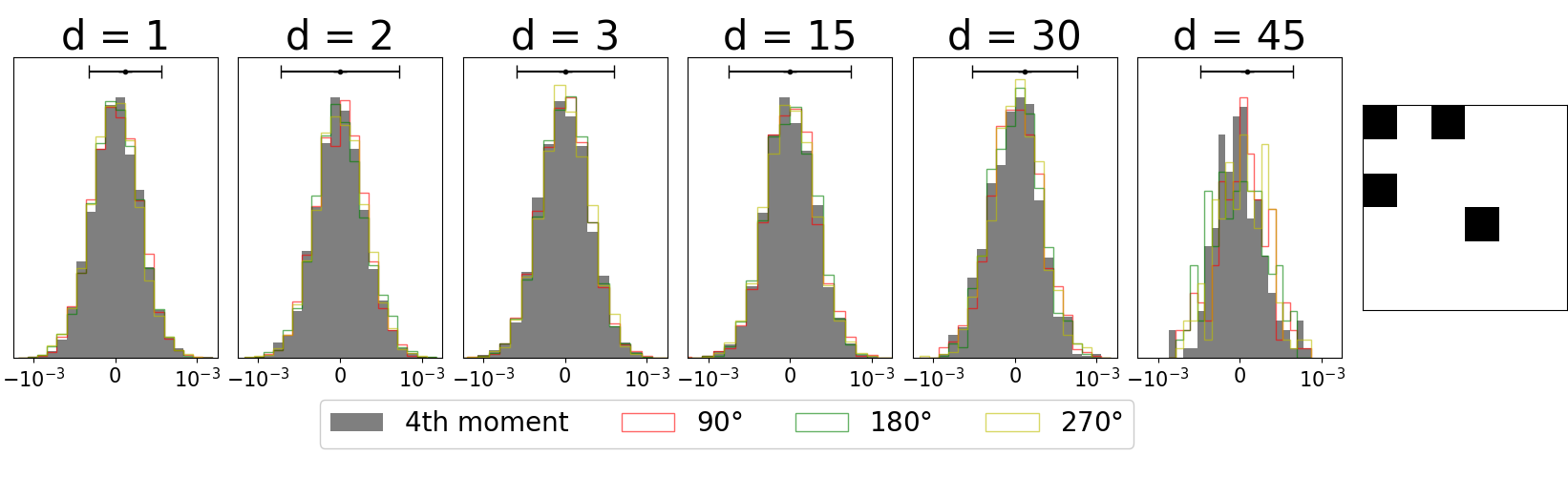}
    \caption{Illustration of statistical isotropy and homogeneity: plotted are histograms of the fourth generalized (since multivariate) central moment of four pixels.  This fourth moment is measured for the non-Gaussian Q-field. The four pixels are arranged in a geometric shape as depicted in the rightmost panel. The shape is shifted and rotated around the field and the fourth moment is histogrammed for each instance. The grey histograms arises from shifting the shape as shown, the red, green and yellow histograms first rotate the shape by 90, 180 or 270 degrees and then shift all over the field. All four histograms agree, which supports statistical isotropy and homogeneity.}
    \label{fig:iso_hom}
\end{figure*}
Standard cosmology demands cosmic fields to be statistically homogeneous and isotropic at the largest scales. We implemented these constraints in Sect.~\ref{sec:isohom}, and verified isotropy and homogeneity as follows.

We compute the 4th multivariate central moment for irregular 4-pointed shapes of pixels. An example of such a 4-pixel shape is depicted in Figure \ref{fig:iso_hom}.  The multivariate 4th central moment is estimated as:
\begin{equation}
    f(x_\alpha, x_\beta, x_\gamma, x_\delta) = \frac{1}{N}\sum_{i=0}^N (x_{\alpha, i} - \overline{x}_\alpha) (x_{\beta, i} - \overline{x}_\beta) (x_{\gamma, i} - \overline{x}_\gamma) (x_{\delta, i} - \overline{x}_\delta) \,,
\end{equation}
where $\alpha, \beta, \gamma, \delta$ denote the four corners of the shape and are held fixed. The overbar denotes the sample mean.

The fixed shape is then shifted across the field both horizontally and vertically, and the positions are enumerated by the index $i$. The histogram of the 4th moment at each point is shown in Fig.~\ref{fig:iso_hom}.  The different thin lined histograms in colour show the same measurement, but with the shape rotated by either $90\deg$, $180\deg$ or $270\deg$. All these different rotations yielding the same result proves isotropy of the field. The 6 different panels show different separations $d$ between the points in our 4-point shape to indicate that the field is homogeneous and isotropic on all scales. The noisiness of the histograms at larger separations is simply because a smaller number of unique shape positions fit into the field.

\section*{Acknowledgements}
It is a pleasure to thank Harry van Zanten and Joris Bierkens for scientific discussions. We thank Simon Portegies Zwart, Erik Deul, Hermen Jan Hupkes and Marco Streng for supporting the acquisition of the new high-memory node for ALICE. 
\vspace{0.5cm}

\bibliographystyle{mnras}
\bibliography{TDist.bib}

\end{document}